\documentstyle[psfig,aps,prl,multicol,epsf,amssymb]{revtex}

\draft
\begin{document}
\title{
Strong field approximation to the relativistic channeling of electrons in the presence of
electromagnetic waves.
}
\author{
Julio San Rom\'an, Luis Plaja and Luis Roso
}
\address{
Departamento de F\'\i sica Aplicada, Universidad de
Salamanca, E-37008 Salamanca, Spain
}
\date{\today}
\maketitle

\begin{abstract}

We present a study of the interaction of a relativistically planar channeled electron with an intense electromagnetic
field. Using a S-Matrix approach in the Strong Field Approximation, it is shown that the crystal periodicity
affects drastically the excitation process, suppressing the possibility of multiphoton absorption except for some
particular cases. This selective excitation opens the possibility to control the dynamics of the channeling process by
means of an external field. Explicit expressions for the S-matrix N-photon excitation rates together with the corresponding
conservation laws are obtained from the relativistic quantum mechanical Dirac equation. 

\end{abstract}


\begin{multicols}{2}

\section{Introduction}

Channeling in crystal lattices occurs when an accelerated charged particle is introduced into a crystalline target at
sufficiently large energy. Depending on the crystal orientation, the particle's trajectory may be aligned
with a major crystal direction and the penetration may reach anomalous depths. Although the possibility of
this effect was already pointed out very early by Stark \cite{stark}, it was demonstrated experimentally 50
years later by Rol {\em et al.} \cite{rol}, when the result of the ion sputtering was found to depend strongly on 
the orientation of the target crystal. After the discovery, the theoretical and experimental work increased rapidly
and extended to the case of channeling of electrons and positrons \cite{gemmell}.

In this paper we will investigate the excitation dynamics of a planar channeled electron under the influence of an
external electromagnetic field. As a main result, we demonstrate that the crystal periodicity introduces a
momentum conservation condition which affects the efficiency of the different N-photon channels of
excitation, leading to the strong suppression of photon absorption in a broad range of situations. The
multiphoton excitation of channeled particles has been already addressed by Avetissian {\em et al.}
\cite{avetissian,avetissian2} by assuming an electromagnetic wave copropagating with the electron, and with a
frequency resonant to the (Doppler-shifted) lower-energy level transitions. In the present case, however, we are
interested in a complementary situation where the channeled electron is excited to a final state lying in the
crystal quasi-continuum. Since the transition is produced by the interaction of the electron with an
external intense optical field, the strong field approximation (SFA)
constitutes a more appropriated procedure in comparison to the discrete level
approach in \cite{avetissian}. 

SFA theories have been developed in the context of ionization of atoms in strong laser fields. Among them,
the so-called Keldish-Faisal-Reiss (KFR) theory
\cite{keldysh,faisal,reiss} is based on the S-matrix approach, where the final
state is approximated by a Volkov state, which describes the evolution of a free electron driven by the electromagnetic
wave. SFA theories describe most of the relevant aspects of the atomic ionization, including multiphoton absorption and
multiphoton excitation above the ionization threshold (ATI). 

Although employed mainly in the atomic and molecular context,
S-matrix SFA approaches can be used in any general situation in which the field interaction energy is comparable with the
energies of the matter system. In fact, for the higher energy
boundstates, the intensity of the field required to promote an electron to the continuum does
not have to be very high, and yet SFA can be used. On the other hand, SFA requires the matter potential to be
approximately constant over the complete interaction time. In our case it suffices with a moderate intensity field,
($10^{12}-10^{13} W/cm^2$), while the crystal stability can be ensured by a sufficiently short pulse (about $100 fs$), which
still enclose enough cycles to ensure the adiabatic limit involved in the theoretical approach. It should be
mention that, in the case of atom ionization by strong field, the adiabatic assumption is correct even
in the case of a few cycles pulse-length, and there is no reason to think that in this channeling case the
thing should be different.

\section{Geometry of the system and description of the unperturbed channeled electron.}
\label{sec:geometry}

\begin{figure}
\psfig{figure=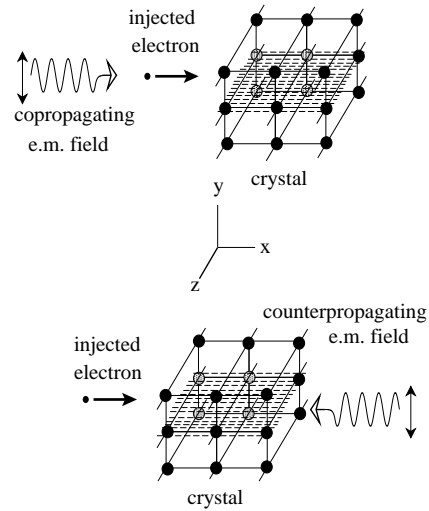,width=8.0cm}
\vspace*{1.00cm}
\caption
{\narrowtext
The system to be studied consists of a relativistic electron channeled
in a crystal and interacting with a copropagating (top picture) or
a counterpropagating (bottom picture) electromagnetic field.
}
\label{geometry}
\end{figure}

Let us consider the interacting geometry depicted in fig. \ref{geometry}. A relativistic electron, with velocity
parallel to the $x-axis$, is introduced in a crystal while interacting with a copropagating or
a counterpropagating electromagnetic wave. For particular crystal orientations, the electron is
confined transversely to trajectories close to the initial injection axis. Our case of planar channeling
occurs when the electron is injected parallel to a crystal plane \cite{gemmell,lindhard}.  We will assume an electromagnetic planewave field, which
propagates in the same direction of the injected electron, and linearly polarized in the $y-axis$, i.e. orthogonal to the
crystal plane.

Since the channeled particle is injected with a relativistic velocity, the crystal potential may be well
approximated by a spatial average over the crystal plane coordinates, as it is done in the so-called continuum model
\cite{gemmell},
\begin{equation}
V(y)= {1 \over {L_x L_z}} \int_{-L_x/2}^{L_x/2}\int_{-L_z/2}^{L_z/2} U(x,y,z) dxdz
\end{equation}

\noindent
where $L_x$ and $L_y$ are the crystal plane dimensions, and $U(x,y,z)$ is the crystal potential. With
this effective potential the electron motion in the crystal channel, driven by the external e.m. field,
will be confined to the polarization plane $xy$. 

In the most general case, the quantum description of the electron's dynamics is described by the Dirac equation
\begin {equation}
\label{dirac}
\left\{ c{\bf \alpha}\cdot \left( {\bf \hat p}- e {\bf A}/c \right) +\beta mc^2 +V(y) \right\} \Phi ({\bf
x})=E_B
\Phi ({\bf x})
\end {equation}

Let us first consider the unperturbed channeling situation, ${\bf A}={\bf 0}$.
Since the averaged potential depends only on the $y$ coordinate, a general positive energy solution of the
 channeled electron can be written as

\begin {equation}
\label {totalsol}
\Phi ({\bf x})= \int dp_y \sqrt{\frac{mc^2}{E_{p_x ,p_y}}}
u^1_{p_x ,p_y} e^{i(p_x x+p_y y)/\hbar} \xi_{p_y}
\end {equation}

\noindent
Where $u^1_{p_x ,p_y}$ is the positive energy solution of a free Dirac electron \cite{bjorken,greiner}, $E_{p_x
,p_y} =\sqrt{c^2 p_x^2 +c^2 p_y^2 +m^2 c^4}$,

\begin {equation}
\label{u1sol}
u^1_{p_x ,p_y} =\sqrt{\frac{E_{p_x ,p_y} +mc^2}{2mc^2}}
\pmatrix{1\cr 0\cr 0\cr \frac{c(p_x +ip_y)}{E_{p_x ,p_y} +mc^2}\cr}
\end {equation}

Introducing (\ref{totalsol}) into (\ref{dirac}), with ${\bf A}={\bf 0}$, one can
obtain the following form for the Dirac equation in momentum space:

\end{multicols}

\vskip 0.1 true in
\line(1,0){450}
\vskip 0.1 true in

\begin {equation}
\label{D_eq_enp}
E_B u^1_{p_x ,p_y} \xi_{p_y} = \{ c\alpha_x p_x +c\alpha_y p_y +\beta mc^2 \}
u^1_{p_x ,p_y} \xi_{p_y} +\int d{p'}_y \sqrt{\frac{E_{p_x ,p_y}}{E_{p_x ,{p'}_y}}}
V_{p_y -{p'}_y} u_{p_x ,{p'}_y} \xi_{{p'}_y}
\end {equation}

\vskip 0.1 true in
\line(1,0){450}
\vskip 0.1 true in

\begin{multicols}{2}

\noindent
Being $V_{p}$ the Fourier transform of the interplanar potential at the spatial frequency  $p/\hbar$.
Due to the nature of the averaged potential, the channeling along a crystal plane is only possible if the
energy of the electron's transversal dynamics is moderate, i.e. non relativistic. In this case, it is justified
to approximate equation (\ref{D_eq_enp}) to second order of
$cp_y/E_{p_x}$.

\begin {equation}
\label{approx_e}
E_{p_x ,p_y} =\sqrt{c^2 p_x^2 +c^2 p_y^2 +m^2 c^4} \approx E_{p_x}
\biggl( 1+ \frac{c^2 p_y^2}{2E_{p_x}^2} \biggr)
\end {equation}

\begin {equation}
\label{approx_raiz}
\sqrt{\frac{E_{p_x ,p_y} +mc^2}{2E_{p_x ,p_y}}} \approx \sqrt{\frac{E_{p_x} +mc^2}{2E_{p_x}}}
\biggl( 1-\frac{c^2 p_y^2}{4E_{p_x}^2}\biggl( \frac{mc^2}{E_{p_x} +mc^2}\biggl) \biggr)
\end {equation}

\begin {equation}
\label{approx_raiz_inv}
\sqrt{\frac{2E_{p_x ,p_y}}{E_{p_x ,p_y} +mc^2}} \approx \sqrt{\frac{2E_{p_x}}{E_{p_x} +mc^2}}
\biggl( 1+\frac{c^2 p_y^2}{4E_{p_x}^2}\biggl( \frac{mc^2}{E_{p_x} +mc^2}\biggl) \biggr)
\end {equation}
Note that, for the field intensities considered here, this approximation will remain equally valid  when
considering the interaction with the external electromagnetic wave.

By introducing eq. (\ref{approx_e}-\ref{approx_raiz_inv}) into (\ref{D_eq_enp}), and since the
scalar potential
$V_{p_y -p_y}$ is a first order term in $cp_y/E_x$, the Dirac equation is reduced to the same identity for
every non-zero component of the spinor $u^1_{p_x ,p_y}$:
\begin {equation}
\label{sch_enp}
E_B \xi_{p_y} = \{ \frac{c^2 p_y^2}{2E_{p_x}} +E_{p_x} \} \xi_{p_y} +
\int d{p'}_y V_{p_y -{p'}_y} \xi_{{p'}_y}
\end {equation}
Computing the inverse Fourier transform in the $y$ coordinate, we finally end with a Schr\"odinger-like equation
\cite{kimball,fusina}
\begin {equation}
\label{sch_eny}
\epsilon_B \xi (y)= \{ \frac{p_y^2}{2\gamma_x m} +V(y) \}\xi (y)
\end {equation}

\noindent
where $E_{p_x} =\gamma_x mc^2$, and $\epsilon_B =E_B -\gamma_x mc^2$ corresponds to the non-relativistic
eigenstate energy.

Several explicit forms of the averaged crystal potential may be found in the
literature \cite{gemmell}. Among them, those derived from the Thomas-Fermi screened two-body potentials have
been widely used \cite{gemmell,lindhard}, and have been fine adjusted by standard
Hartree-Fock many-body calculations \cite{klein}. From the theoretical point of view, model potentials are
more convenient, since they allow for further analytical work while keeping the essential features of the
interaction.  For
instance, $V(y)=-4V_0 y^2/d^2$ was proposed by Avetissian {\it et al.}
\cite{avetissian} in the context of computation of the multiphoton transitions between bound states of the
continuum potential, for a positron interacting with a strong electromagnetic
wave. The form $V(y)=-V_0\cosh^{-2} (y/b)$ has also been used to study
the possible amplification of x-ray channeling radiation \cite{avetissian2}. This later form has a better
resemblance with the averaged Thomas-Fermi potentials while still allowing for an analytical diagonalization and,
therefore, we shall use this potential for our calculations. The transverse energy spectrum for this case can be cast in
the following form \cite{avetissian2}:

\begin {equation}
\label{E_spectrum}
{\epsilon_B}_n =-\frac{\hbar^2}{2b^2 m\gamma_x} (s-n)^2
\end {equation}

\noindent
where $n$ can be $0,1,...,[s]$, being
$s=-\frac{1}{2}+\sqrt{\frac{1}{4}+\frac{2b^2 m\gamma_x V_0}{\hbar^2}}$. Note that, as a result of the spatial
averaging, the scattering with the continuum potential affects only to the transverse dynamics of the
channeled particle, while the very large longitudinal momentum remains unaffected.

%\pagebreak

\section {S matrix description of a channeled particle interacting with an electromagnetic wave.}

Let us now add the electromagnetic excitation to the problem. As it is well known, eq.
(\ref{dirac}) does not accept analytical solutions for a space-time dependent vector potential. In such situations,
the S-matrix approach offers a standard procedure to find approximated solutions \cite{wheeler}, used specially in
quantum field theory \cite{bjorken,schweber,itzykson} and scattering
\cite{roman}.

\subsection{The general relativistic case}

The relativistic SFA S-matrix theory for the Dirac electron in an atom can be found in
\cite{reiss80,reiss90,reissPQE}. The general expression for the transition amplitude
using time-reversed S matrix theory has the following form
\begin {equation}
\label{smatrix1}
S_{fi} = \lim_{t\rightarrow\infty} \langle{\Psi_f^{(-)} \vert \Phi_i}\rangle
\end {equation}

\noindent
Although mainly used in the
strong field ionization of atoms and molecules, this approach is quite general and can be exported to any other
system, provided its eigenstates can be found analytically. To our knowledge, however, this is the first time
that it is applied to the relativistic channeled electron in interaction with an electromagnetic wave. In the present
case,
$\Phi_i$ corresponds to the unperturbed channeled electron state discussed in sec. \ref{sec:geometry}, and
$\Psi_f^{(-)}$ is an arbitrary final state, solution of the complete equation (\ref{dirac}). Since this exact
solution is not available, the success of the S-matrix approach consist in finding a
suitable approximation. In the strong field approximation, the interaction with electromagnetic field is assumed to be
the relevant for the final state, therefore $\Psi_f^{(-)}$ is approximated in terms of the Volkov states,
$\Psi_V ({\bf x},t)$
\cite{wolkow,bknr}. These wavefunctions are solutions of eq. (\ref{dirac}) for $V(y)=0$ and ${\bf A}({\bf x},t) \neq {\bf
0}$, and describe a free electron in the presence of an electromagnetic field.  The form of these states
for a laser field pulse of arbitrary form is \cite{bknr}:
\begin {equation}
\label{genvolkov}
\Psi_V^{(-)} (x)=\sqrt{\frac{mc^2}{(2\pi )^3 E}}
\biggr( 1+\epsilon^r \frac{e{s\llap/}{A\llap/}}{2c(s\cdot p)} \biggr)
u^r_{\bf p} e^{iS}
\end {equation}

\noindent
being $\epsilon^{1,2} =1$, $\epsilon^{3,4} =-1$, and

\begin{equation}
\label{eq:volkov-phase}
S = -\epsilon^r \frac{p\cdot x}{\hbar} +
\int^{\infty}_{s\cdot x} \biggr[ \frac{e(p\cdot A(\varphi '))}{\hbar c(s\cdot p)}
-\epsilon^r \frac{e^2 A^2 (\varphi ')}{2\hbar c^2 (s\cdot p)}\biggl] d\varphi '
\end{equation}

The S-matrix approach takes as a starting point the following exact
relation
\begin {equation}
\label{expansion}
\Psi_f (x)=\Psi_V (x)+\int d^4 x' G_V (x,x')\gamma^0 V(x')\Psi_f (x')
\end {equation}

\noindent
where $G_V (x,x')$ is the Volkov Green's function \cite{reiss66}, and keeps the lowest order term in powers of
$V(x)$. The transition amplitude obtained is \cite{reiss90,reissPQE}:

$$
(S-1)_{fi}^{SFA} = -\frac{i}{\hbar c} \int{d^4 x {\overline{\Psi}_V^{(-)}} eA_\mu \gamma^\mu \Phi_i}
$$
\begin{equation}
\label{int4}
=-\frac{i}{\hbar c} \sqrt{\frac{mc^2}{(2\pi )^3 E}}
\int d^4 x  e^{-iS}\overline{u}^r_{\bf p}
\biggr( e{A\llap/}-\epsilon^r \frac{e^2 (A\cdot A){s\llap/}}{2c(s\cdot p)} \biggr) \Phi_i
\end {equation}

\noindent
where we have used (\ref{genvolkov}) and we have assumed the transverse character of the e.m. field, $s\cdot
A=-{\bf s}\cdot{\bf A}=0$. Separating the time and the space integrals in equation (\ref{int4}), the transition amplitude
takes the form:

\end{multicols}

\vskip 0.1 true in
\line(1,0){450}
\vskip 0.1 true in

\begin {equation}
\label{tamplitude}
(S-1)_{fi}^{SFA} =-i \sqrt{\frac{mc^2}{(2\pi )^3 E}} e\overline{u}^r_{\bf p} \int{dt {\Im\llap/}_1 (t)}+
i \sqrt{\frac{mc^2}{(2\pi )^3 E}} \frac{e^2 \epsilon^r}{2(s\cdot p)}
\overline{u}^r_{\bf p} {s\llap/} \int{dt \Im_2 (t)}
\end {equation}
being ${\Im\llap/}_1 (t)$ and $\Im_2 (t)$:
\begin {eqnarray}
\label{i1}
{\Im\llap/}_1 (t)&=&\gamma^\mu \Im_{1,\mu} (t)=\frac{\gamma^\mu}{\hbar}
\int d{\bf x} e^{-iS} A_\mu e^{-i\frac{E_B t}{\hbar}} \Phi_i \\
\label{i2}
\Im_2 (t)&=&\frac{1}{\hbar c} \int d{\bf x} e^{-iS} (A\cdot A) e^{-i\frac{E_B t}{\hbar}} \Phi_i =
-\frac{1}{\hbar c} \int d{\bf x} e^{-iS} \vert {\bf A}\vert^2 e^{-i\frac{E_B t}{\hbar}} \Phi_i
\end {eqnarray}

\vskip 0.1 true in
\line(1,0){450}
\vskip 0.1 true in

\begin{multicols}{2}

Once the transition amplitude $(S-1)_{fi}^{SFA}$ is known, the total transition rate can be computed as:
\begin {equation}
\label{ttr}
W=\int \frac{V d{\bf p}}{(2\pi )^3}w
\end {equation}
where $V$ is the normalization volume and $w$ is the transition probability per unit
of time, which is defined as

\begin {equation}
\label{tpput}
w=\lim_{t\rightarrow \infty} \frac{1}{t} \vert (S-1)_{fi}^{SFA} \vert^2
\end {equation}

Another important magnitude is the transition
rate per unit of solid angle, which has the form:

\begin {equation}
\label{trpusa}
\frac{dW}{d\Omega}=\frac{V}{(2\pi )^3} \int wp^2 dp
\end {equation}

It should be mention that SFA S-matrix theory cannot be considered as a perturbation series in powers of $V(x)$
since the initial state $\phi_i$ is an eigenstate of the potential itself including all the properties of the
crystal. The presence of this wavefunction produces a new behavior in the scattering section, introducing all the
differences with the atom ionization or the simple Compton scattering.

\subsection{Application to the case of a monochromatic and linearly polarized laser field}

Lets us focus our attention to the geometry depicted in fig. \ref{geometry}, where the electromagnetic field
can be a copropagating or counterpropagating linear polarized planewave of frequency $\omega$,
${\bf A}(\varphi)=A_j (\varphi){\bf e}_j =A_0 \cos (\omega t-{\bf k}\cdot{\bf x}) {\bf e}_j$.
where ${\bf e}_j$ is the field's polarization vector.

The phase factor of the Volkov function (\ref{eq:volkov-phase}) now reads as

\end{multicols}

\vskip 0.1 true in
\line(1,0){450}
\vskip 0.1 true in

\begin {equation}
S = -\epsilon^r \biggl( \frac{p^\mu}{\hbar}+\frac{e^2 A_0^2 k^\mu}{4\hbar c^2 (k\cdot p)} \biggr)x_\mu
+\frac{ep_j A_0}{\hbar c (k\cdot p)}\sin (k\cdot x)-
\epsilon^r \frac{e^2 A_0^2}{8\hbar c^2 (k\hbar p)}\sin  (2(k\cdot x))
\end {equation}

\noindent
The resulting exponential factor can be expanded as a series of Bessel functions

\begin {equation}
\label{bessel}
e^{-iS} = e^{i\epsilon^r \bigl[ \frac{p^\mu}{\hbar}+\frac{e^2 A_0^2 k^\mu}{4\hbar c^2 (k\cdot p)}\bigr]x_\mu}
\sum_{N,n=-\infty}^{+\infty}{J_{N+2n}(\eta) J_{n}(\xi) e^{-iN(k\cdot x)}}
\end {equation}

\noindent
where $\eta$ and $\xi$ are factors which only depend on the momentum of the Volkov
function, and the frequency and amplitude of the laser field:

\begin {equation}
\eta =\frac{ep_j A_0}{\hbar c(k\cdot p)}
\quad \quad
\xi = \frac{\epsilon^r e^2 A^2_0 }{8\hbar c^2 (k\cdot p)}
\end {equation}

Substituting (\ref{bessel}) in eqs. (\ref{i1}) and (\ref{i2}):

\begin{eqnarray}
{\Im\llap/}_1 (t) &= &-\gamma^j \Im_{1,j} (t)=
-\frac{\gamma^j}{\hbar} \int{d{\bf x} e^{-iS} A_j (\varphi) \Phi_i ({\bf x})} \nonumber \\
\label{i1sfa}
& =& -\frac{\gamma^j A_0}{\hbar} \sum_{N,n=-\infty}^{+\infty} \frac{N+2n}{\eta}J_{N+2n}(\eta)
J_{n}(\xi)e^{i\frac{(\hbar \omega_N -E_B )t}{\hbar}} \tilde{\Phi}_i ({\bf q}_N )
\end {eqnarray}

\noindent
and
\begin{eqnarray}
\Im_2 (t)&=&-\frac{1}{\hbar c} \int d{\bf x} e^{-iS} \vert {\bf A}\vert^2
e^{-i\frac{E_B t}{\hbar}} \Phi_i \nonumber \\
&=&-\frac{A_0^2}{\hbar c} \sum_{N,n=-\infty}^{+\infty}
\biggl( \frac{(N+2n)^2}{\eta^2}J_{N+2n} (\eta) +\frac{1}{2\eta}(J_{N+2n+1} (\eta)-J_{N+2n-1} (\eta))\biggr)
\nonumber \\
\label {i2sfa}
& & \times J_n (\xi) e^{i\frac{(\hbar \omega_N -E_B )t}{\hbar}} \tilde{\Phi}_i ({\bf q}_N )
\end {eqnarray}
Where ${\bf q}_N$, $\omega_N$ and $\tilde{\Phi}_i ({\bf q}_N )$ are defined as:

\begin {eqnarray}
\label{qn}
{\bf q}_N &=& \frac{\epsilon^r {\bf p}}{\hbar}+
\biggl[ \frac{\epsilon^r e^2 A^2_0}{4\hbar c^2 (k\cdot p)} -N \biggr]{\bf k} \\
\label{wn}
\omega_N &=& \frac{\epsilon^r E}{\hbar}+
\biggl[ \frac{\epsilon^r e^2 A^2_0}{4\hbar c^2 (k\cdot p)} -N \biggr]\omega \\
\label{fitilde}
\tilde{\Phi}_i ({\bf q}_N )&=&\int{e^{-i{\bf q}_N \cdot{\bf x}} \Phi_i ({\bf x}) d{\bf x}}
\end {eqnarray}

The time integrals appearing in (\ref{tamplitude}) can now be calculated as
\begin{eqnarray}
\gamma^j \int_{t_0}^{t}{\Im_{1,j} (\tau) d\tau} &=&\frac{\gamma^j A_0}{\hbar} \sum_{N,n=-\infty}^{+\infty}
\frac{N+2n}{\eta}J_{N+2n}(\eta) J_{n}(\xi) \tilde{\Phi}_i ({\bf q}_N )
\nonumber \\
\label{tint1}
& & \times e^{i(\hbar \omega_N -E_B )(t+t_0 )/2\hbar}
\frac{\sin{(\hbar \omega_N -E_B )T/2\hbar}}{(\hbar \omega_N -E_B )/2\hbar}
\end {eqnarray}

\noindent
and
\begin{eqnarray}
\int_{t_0}^{t}{\Im_2 (\tau)d\tau}&=&
-\frac{A_0^2}{\hbar c} \sum_{N,n=-\infty}^{+\infty}
\biggl( \frac{(N+2n)^2}{\eta^2}J_{N+2n} (\eta) +\frac{1}{2\eta}(J_{N+2n+1} (\eta)-J_{N+2n-1} (\eta))\biggr)
 \nonumber \\
\label{tint2}
& & \times J_n (\xi) \tilde{\Phi}_i ({\bf q}_N )
e^{i(\hbar \omega_N -E_B )(t+t_0 )/2\hbar}
\frac{\sin{(\hbar \omega_N -E_B )T/2\hbar}}{(\hbar \omega_N -E_B )/2\hbar}
\end {eqnarray}
being $T=t-t_0$

To compute the rate of excitation, we should use this two equations together with eq. (\ref{tamplitude}) to
calculate
\begin{equation}
\vert (S-1)_{fi}^{SFA}\vert^2 = T_1 +T_2 +T_3 +T_4
\end {equation}
with
\begin {eqnarray}
T_1 &=& \frac{mc^2}{(2\pi)^3 E}e^2 \Bigl[ \int{dt \Im_{1,j} (t)} \Bigr]^{+}{\gamma^j}^+
\gamma^0 u^r_{\bf p}{{u}^r_{\bf p}}^+ \gamma^0 \gamma^j \int{dt \Im_{1,j} (t)}
\\
T_2 &=&\frac{mc^2}{(2\pi)^3 E}\frac{\epsilon^r e^3}{2(k\cdot p)}
\Bigl[ \int{dt \Im_{1,j} (t)} \Bigr]^{+}{\gamma^j}^+
\gamma^0 u^r_{\bf p}{{u}^r_{\bf p}}^+ \gamma^0 {k\llap/} \int{dt \Im_2 (t)}
\\
T_3&=&T_2^*
\\
T_4 &=&\frac{mc^2}{(2\pi)^3 E}\biggl(\frac{\epsilon^r e^2}{2(k\cdot p)}\biggr)^2
\Bigl[ \int{dt \Im_2 (t)} \Bigr]^{+}{\gamma^\mu}^+ k^\mu
\gamma^0 u^r_{\bf p}{{u}^r_{\bf p}}^+ \gamma^0 {k\llap/} \int{dt \Im_2 (t)}
\end {eqnarray}

Substituting the time integrals, \ref{tint1} and \ref{tint2}, in each term, we may calculate
the transition probability per unit of time as:

\begin {equation}
\label{c_tpput}
w=\lim_{t\rightarrow \infty} \frac{1}{t} \vert (S-1)_{fi}^{SFA} \vert^2 =
\lim_{t\rightarrow \infty} \frac{T_1 +T_2 +T_3 +T_4}{t}=t_1 +t_2 +t_3 +t_4
\end {equation}

\noindent
where
\begin{eqnarray}
t_1 &=& \frac{mc^2}{8\pi^2 E}\frac{e^2 A_0^2}{\hbar^2}
\sum_{N=-\infty}^{+\infty} \biggl[ \sum_{n=-\infty}^{+\infty} \frac{(N+2n)}{\eta}
J_{N+2n} (\eta) J_{n} (\xi) \biggr]^2 \delta(\hbar \omega_N -E_B )
\nonumber \\
\label{tlimit1}
&  & \times \tilde{\Phi}_i^+ ({\bf q}_N )
{\gamma^j}^+ \gamma^0 u^r_{\bf p}{{u}^r_{\bf p}}^+ \gamma^0 \gamma^j \tilde{\Phi}_i ({\bf q}_N )
\\
t_2 &=& -\frac{mc^2}{8\pi^2 E}\frac{\epsilon^r e^3 A_0^3}{2\hbar^2 c (k\cdot p)}
\sum_{N=-\infty}^{+\infty} \sum_{n,n'=-\infty}^{+\infty}
\biggl( \frac{(N+2n')^2}{\eta^2}J_{N+2n'} (\eta) +\frac{1}{2\eta} (J_{N+2n'+1} (\eta)-J_{N+2n'-1} (\eta)) \biggr)
J_{n'} (\xi)
\nonumber \\
\label {tlimit2}
& & \times \frac{(N+2n)}{\eta} J_{N+2n} (\eta) J_{n} (\xi)
\delta (\hbar \omega_N -E_B ) \tilde{\Phi}_i^+ ({\bf q}_N )
{\gamma^j}^+ \gamma^0 u^r_{\bf p}{{u}^r_{\bf p}}^+ \gamma^0 {k\llap/}
\tilde{\Phi}_i ({\bf q}_N )
\\
t_3 &=& t_2^*
\label {tlimit3}
\\
t_4 &=& \frac{mc^2}{8\pi^2 E}\biggl(\frac{\epsilon^r e^2 A_0^2}{2\hbar c (k\cdot p)}\biggr)^2
\sum_{N=-\infty}^{+\infty}
\biggl[ \sum_{n=-\infty}^{+\infty} \biggl( \frac{(N+2n)^2}{\eta^2} J_{N+2n} (\eta) +\frac{1}{2\eta} (J_{N+2n+1} (\eta)-J_{N+2n-1} (\eta)) \biggr) \biggr]^2
\nonumber \\
\label{tlimit4}
& & \times \delta (\hbar \omega_N -E_B ) \tilde{\Phi}_i^+ ({\bf q}_N )
{k\llap/}^+ \gamma^0 u^r_{\bf p}{{u}^r_{\bf p}}^+ \gamma^0 {k\llap/} \tilde{\Phi}_i ({\bf q}_N )
\end {eqnarray}

Finally, the excitation rate is

\begin {equation}
\label{result}
w=\frac{mc^2}{8\pi^2 E}\biggl( \frac{eA_0}{\hbar} \biggr)^2 \sum_{N=-\infty}^{+\infty}
\biggl( S_1^2 \Delta_1 -\frac{\epsilon^r e A_0}{c(k\cdot p)}
S_2 Re(\Delta_2 ) +\biggl( \frac{e A_0}{2c(k\cdot p)}\biggr)^2 S_4^2 \Delta_4 \biggr)
\delta (\hbar \omega_N -E_B )
\end {equation}

Where we have defined:

\begin {eqnarray}
S_1 &=& \sum_{n=-\infty}^{+\infty} \frac{(N+2n)}{\eta}
J_{N+2n} (\eta) J_{n} (\xi)
\\
S_2 &=& \sum_{n,n'=-\infty}^{+\infty}
\biggl( \frac{(N+2n')^2}{\eta^2}J_{N+2n'} (\eta) +
\frac{1}{2\eta} (J_{N+2n'+1} (\eta)-J_{N+2n'-1} (\eta)) \biggr)
J_{n'} (\xi)
\nonumber \\
& & \times \frac{(N+2n)}{\eta} J_{N+2n} (\eta) J_{n} (\xi)
\\
S_4 &=& \sum_{n=-\infty}^{+\infty} \biggl( \frac{(N+2n)^2}{\eta^2} J_{N+2n} (\eta) +
\frac{1}{2\eta} (J_{N+2n+1} (\eta)-J_{N+2n-1} (\eta)) \biggr)  J_{n} (\xi)
\\
\Delta_1 &=& \tilde{\Phi}_i^+ ({\bf q}_N )
{\gamma^j}^+ \gamma^0 u^r_{\bf p}{{u}^r_{\bf p}}^+ \gamma^0 \gamma^j \tilde{\Phi}_i ({\bf q}_N )
\\
\Delta_2 &=& \tilde{\Phi}_i^+ ({\bf q}_N )
{\gamma^j}^+ \gamma^0 u^r_{\bf p}{{u}^r_{\bf p}}^+ \gamma^0 {k\llap/} \tilde{\Phi}_i ({\bf q}_N )
\\
\Delta_4 &=& \tilde{\Phi}_i^+ ({\bf q}_N )
{k\llap/}^+ \gamma^0 u^r_{\bf p}{{u}^r_{\bf p}}^+ \gamma^0 {k\llap/} \tilde{\Phi}_i ({\bf q}_N )
\end {eqnarray}

\vskip 0.1 true in
\line(1,0){450}
\vskip 0.1 true in

\begin{multicols}{2}

\section {Conservation laws and closing of excitation channels.}

As expressed in eq. (\ref{result}), the transition probability is a function of the initial momentum-space
probability amplitude, $\tilde{\Phi}_i ({\bf q}_N )$. Let us now assume injected electron of positive energy
with a wavefunction of the form (\ref{totalsol}), therefore $\epsilon^r =+1$. From
the delta function in eq. (\ref{result}),  we obtain the following energy conservation relation:
\begin {equation}
\label{E_conserv}
E+\biggl[ \frac{e^2 A^2_0}{4\hbar c^2 (k\cdot p)} -N \biggr]\hbar \omega=E_B
\end {equation}

\begin{figure}
\psfig{figure=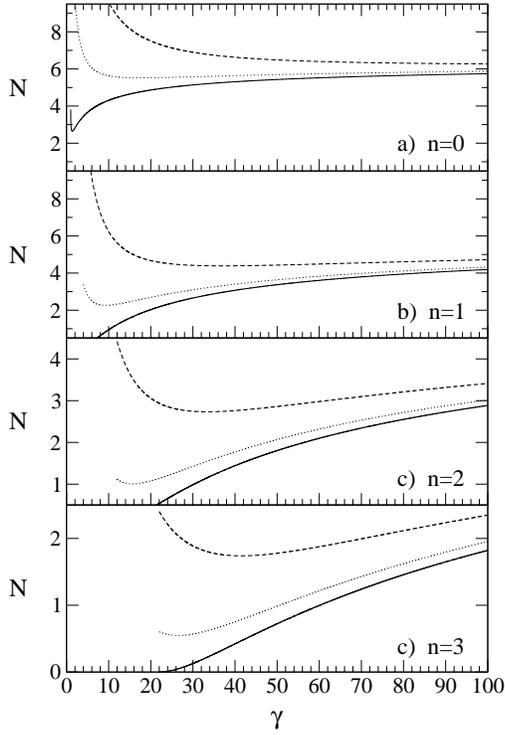,width=8.0cm}
\vspace*{1.00cm}
\caption
{\narrowtext
This figures shows the values of $(\gamma_x )_i$ needed to open
a specific N-photon excitation channel for the lowest orders of transverse
momentum transferred $\tilde{n}$. All the pictures correspond to an electron channeled
along the $(110)$ plane of silicon, driven by a counterpropagating linear
polarized TiSa laser, $\lambda=800$ nm, of $3.51\times 10^{12} W/cm^2$. The
different pictures represent distinct initial bound states of the
interplanar potential. The top one corresponds to the ground state
case ($n=0$), the second picture to the first excited state ($n=1$) and so on. The
continuous, dotted and dashed lines represent the excitation process
with zero, one and two quanta of transverse momentum transferred
($\tilde{n}=0,\pm 1,\pm 2$) respectively.
}
\label{counterpropagating}
\end{figure}

\noindent
On the other hand, an additional conservation law  relates the momentum of the final and initial states,
${\bf p}$ and
${\bf p}_i$ respectively, in
(\ref{qn}). For a positive energy electron, this reads as
\begin {equation}
\label{p_conserv}
{\bf p}_i =\frac{{\bf p}}{\hbar}+
\biggl[ \frac{e^2 A^2_0}{4\hbar c^2 (k\cdot p)} -N \biggr]{\bf k}
\end {equation}
Since the electromagnetic field propagates along the $x$-axis, this condition may be splitted into two
parts
\begin {eqnarray}
\label{p_conservx}
(p_x )_i &=& p_x \pm
\biggl[ \frac{ e^2 A^2_0}{4\hbar c^2 (k\cdot p)} -N \biggr]\hbar k
\\
\label{p_conservy}
(p_y )_i &=& p_y
\end{eqnarray}

\noindent
with $(k\cdot p)=k(\frac{E}{c} \mp p_x )$, and where $k=|{\bf k}|$, the top sign refers to a field
copropagating with the electron, and the bottom to the counterpropagating case. These three equations, (\ref{E_conserv}),
(\ref{p_conservx}) and (\ref{p_conservy}), describe the energy and momentum changes
due to the stimulated absorption or emission of $N$ photons.  Combining these with the energy expression for
the final state, $E=\sqrt{c^2 p_x^2 +c^2 p_y^2 +m^2 c^4}$, we obtain a closed formula for the energy
conservation of the multiphoton process, in terms of the initial momentum and the field parameters
\begin {equation}
\label{simplify}
N\hbar \omega (1 \mp (\beta_x )_i )=\frac{c^2 (p_x )_i^2 +c^2 (p_y )_i^2 +m^2 c^4 -E_B^2 +{{A_0^2 e^2}\over{2}}}
{2\gamma_B mc^2}
\end {equation}

\noindent
where we have defined the initial energy of the electron as
$E_B =\gamma_B mc^2$, and  the initial relativistic velocity factor,
$(\beta_x )_i =(p_x )_i/\gamma_B mc$. Since $E_B \approx\epsilon_B +(\gamma_x )_i mc^2$, with $(\gamma_x
)_i=\sqrt{1/(1-(\beta_x )_i^2)}$ we have

\begin {equation}
\label{relation}
N\hbar \omega \gamma_B (1 \mp (\beta_x )_i )\approx
\frac{(p_y )_i^2}{2m} -(\gamma_x )_i \epsilon_B +\frac{A_0^2 e^2}{4m c^2}
\end {equation}
The interpretation of this energy conservation relation is straight-forward if we take as reference system a
frame propagating with the electron, with its initial velocity $(p_x)_i/(\gamma_x )_i m$. The frequency
$\omega'=\omega \gamma_B \left( 1 \mp (\beta_x )_i \right)$ corresponds to the Doppler-shifted electromagnetic
wave, and $(\gamma_x )_i \epsilon_B$ is the result of the Lorentz transform of the bound state energy
(\ref{E_spectrum}) from the laboratory to the moving frame. Equation (\ref{relation}) states the resonance
condition for a (Doppler-shifted) $N$-photon transition from a bound state to a state lying in the crystal
quasi-continuum of momentum given by eqs. (\ref{p_conservx}) and (\ref{p_conservy}). This is, in essence, the
crystal equivalent to the ionization of atoms by intense fields. Note, however, that in the atom case a
ionization channel for any photon number $N$ is always possible, since the initial state is distributed
continuously over the momentum space and, therefore, a non-zero transition probability exists for any
$(p_y)_i$ which fulfills the condition (\ref{relation}). This is not the case for the channeled
electron, since the crystal plane periodicity forces a discretization of the electron states in the
transverse coordinate of the momentum space $p_y =\tilde{n} \frac{2\pi h}{d_p}=\tilde{n} \Delta p_y$,
being $\tilde{n}$ an integer and $d_p$ the interplanar distance. As a consequence, in the general case the $N$-photon
channel of excitation should be strongly suppressed, except in those particular cases in which
\begin {equation}
\label{relation3}
N+\frac{(\gamma_x )_i \epsilon_B}{\hbar \omega'} \approx \frac{1}{\hbar \omega'}
\biggl( \frac{\tilde{n}^2 {\Delta p_y}^2}{2m}+\frac{A_0^2 e^2}{4m c^2}\biggr)
\end {equation}
holds for $N$ and $\tilde{n}$ as integer numbers. As a consequence, this opens the possibility of selective excitation of
channeled electrons in terms of their initial velocity, or permits its control through the variation of the
electromagnetic field parameters. Figure \ref{counterpropagating}a-d show the possible $N$-photon channel excitations as a
function of the initial electron's energy and for the lowest orders of transverse momentum transferred $\tilde{n}$.
Each plot shows the result for a different initial channeling bound-state. We assume planar channeling along
the $(110)$ plane of Si by selecting the potential parameters $V_0 =20.4$ eV and $b=0.03$ nm reproducing
\cite{berman}, and a counterpropagating TiSa laser of $3.51\times 10^{12} W/cm^2$ ($\lambda \simeq 800
nm$). Note that the number of photons $N$ should be an integer quantity, therefore the figure shows clearly
that, except for very particular choices of the electron's initial energy, the excitation channels are closed.

\begin{figure}
\psfig{figure=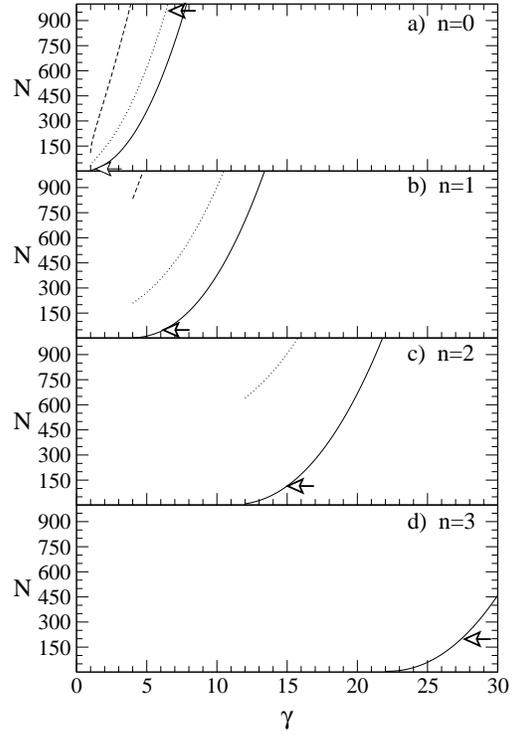,width=8.0cm}
\vspace*{1.00cm}
\caption
{\narrowtext
The same situation as in figure \ref{counterpropagating} but with the external
electromagnetic field copropagating with the channeled electron. The arrows represent
the $N_{Limit}$ for each transition line. The thinner arrow in figure (a) must be only considered as a qualitative
estimation (see text). The transition lines without arrow mean that the $N_{Limit}$
occurs for parameters beyond the plotted region.
}
\label{propagating}
\end{figure}

The same figure can be done for the case of a copropagating electromagnetic field. Figure
\ref{propagating}a-d show again the possible $N$-photon channel excitations as a function of the
initial electron's energy, for the lowest orders of transverse momentum transferred assuming the
same crystal and laser parameters as in figure \ref{counterpropagating}. An important
increase of the number of photons needed to excite the electron, attributable to the Doppler redshift,
can be observed. Under this circumstances, even when the energy and momentum constrains are fulfilled, the
process may involve a very small transition probability due to the high number of photons needed.
To give an idea of the order of this probability one can make use of the asymptotic expansion of the Bessel
functions for large orders, $J_n (x) \approx \frac{1}{\sqrt{2\pi n}}\Bigl( \frac{n_e x}{2n}\Bigr)^n$,
\cite{abramowitz}, being $n_e =\lim_{n\rightarrow \infty} \Bigl( 1+\frac{1}{n}\Bigr)^n$, to calculate
a limit of the number of photons above which the transition probability will be negligible. The criterion
to be used here will be to consider negligible the Bessel function when $\frac{n_e |x|}{2n_{Limit}}\leq 0.1$.
Applying it to our case one obtain the limit of the number of photons for each transverse momentum
transferred as a function of the final energy of the electron:
\begin {equation}
\label{Nlimit}
N_{Limit} =\frac{5n_e |e|A_0}{mc\hbar \omega'} \Bigl( |\tilde{n}| \Delta p_y +\frac{|e|A_0}{2c}\Bigr)
\end{equation}
where $\omega' =\omega \gamma (1-\beta_x)$. Assuming that the electron finishes in a state of the
crystal quasi-continuum and that the energy along the axial direction do not change
significantly during the evolution, one can approximate $\gamma \approx (\gamma_x )_i$ and
$\beta_x \approx (\beta_x )_i$. Consequently $\omega'$, and therefore $N_{Limit}$, can be
expressed as a function of the initial electron's energy. Figure \ref{propagating}a-d show
with arrows the point when the number of photons required for the excitation process surpass
the $N_{Limit}$. For energies above this point, the transition probability reduces drastically and we can consider that
no excitation takes place, even though the energy and momentum conservation relations may be fulfilled. It should be
pointed out that eq. (\ref{Nlimit}) is only valid for the case of large orders in the Bessel functions. This means that it
should be taken qualitatively in all cases in which $N_{Limit}$ is a small quantity, as for instance in figure
\ref{propagating}a for the case $\tilde{n}=0$. Note also that those cases in which the arrow is not shown correspond to
$N_{Limit}$ outside the plotting region, i.e. the $N$-photon excitation is possible along the complete plotted line.

Finally, let us remark the fact that the photon excitation number is greater than in the counterpropagating case increases
the sensitivity of the channel process to the selective excitation in terms of the laser parameters.

%\pagebreak
\section {Conclusions}

We have computed the explicit forms of the S-matrix transition probabilities for the N-photon absorption of a
relativistic electron channeled along a crystal plane. In contrast to previous works, we consider the
interaction with an intense electromagnetic wave, generated externally, which may excite the electron to
high-energy states lying in the crystal quasi-continuum. Due to the crystal periodicity, we show that the energy
and momentum conservation equations constraint strongly this excitation process, suppressing the multiphoton
absorption except for some particular cases. Under these circumstances, the selection of a single multiphoton
channel of excitation is feasible by an adequate choice of the external laser parameters, opening a
broad range of possibilities for the coherent control of the channel electron's dynamics. The case of
an electromagnetic field copropagating with the injected electron is also studied showing an important
increase of the number of photons needed to excite the electron due to the Doppler redshift. For
this case, we give an estimation of the maximum photon number for which the excitation process is not negligible.
The selective excitation in the copropagating case is found to be more sensitive to the electron's energy and the
transverse momentum transferred in the transition than in the counterpropagating one.

\section {Acknowledgments}

We thank enlightening discussions with Professor F.H.M. Faisal. This work has been
supported by the Spanish Direcci\'on General de Ense\~nanza Superior e Investigaci\'on
Cient\'\i fica (grant PB98-0268), and the Junta de Castilla y Le\'on in collaboration
with the European Union, F.S.E. (grant SA044/01).

%\newpage

%\newpage

\end{multicols}

\end{document}